\documentclass[superscriptaddress,amsmath,amssymb, aps, pra, twocolumn]{revtex4-2}

\usepackage{graphicx}
\usepackage{dcolumn}
\usepackage{bm}
\usepackage[colorlinks]{hyperref}
\usepackage[T1]{fontenc}
\usepackage[dvipsnames]{xcolor}
\usepackage{comment}

\begin{document}

\title{Induced on-demand revival in coined quantum walks on infinite $d$-dimensional lattices}

\author{M. N. Jayakody}
\affiliation{Faculty of Engineering and the Institute of Nanotechnology and Advanced Materials, Bar-Ilan University, Ramat Gan 5290002, Israel}

\author{I. L. Paiva}
\affiliation{Faculty of Engineering and the Institute of Nanotechnology and Advanced Materials, Bar-Ilan University, Ramat Gan 5290002, Israel}

\author{A. Nanayakkara}
\affiliation{National Institute of Fundamental Studies, Hanthana Road, Kandy 20000, Sri Lanka}

\author{E. Cohen}
\affiliation{Faculty of Engineering and the Institute of Nanotechnology and Advanced Materials, Bar-Ilan University, Ramat Gan 5290002, Israel}

\begin{abstract}
The study of recurrences and revivals in quantum systems has attracted a great deal of interest because of its importance in the control of quantum systems and its potential use in developing new technologies. In this work, we introduce a protocol to induce full-state revivals in a huge class of quantum walks on a $d$-dimensional lattice governed by a $c$-dimensional coin system. The protocol requires two repeated interventions in the coin degree of freedom. We also present a characterization of the walks that admit such a protocol. Moreover, we modify the quantity known as P\'olya number, typically used in the study of recurrences in classical random walks and quantum walks, to create a witness of the first revival of the walk.
\end{abstract}

\maketitle

\section{Introduction}

Quantum walks (QWs) were introduced as quantum analogs of classical random walks \cite{aharonov1993quantum}. Similarly to their classical counterpart, they have been an important framework for theoretical and practical understanding of quantum algorithms \cite{kempe2003quantum, shenvi2003quantum, childs2003exponential, berry2010quantum} and quantum computing \cite{childs2009universal, childs2013universal}. They have also been used in the modeling of transport in biological systems \cite{oliveira2006decoherence, hoyer2010limits, lloyd2011quantum} and physical phenomena, such as Anderson localization \cite{schreiber2011decoherence, wojcik2012trapping, zhang2014one, crespi2013anderson, xue2014trapping} and topological phases \cite{kitagawa2010exploring, kitagawa2012observation}.

Recurrences have been a subject of extensive study in the QW literature \cite{bouwmeester1999optical, wojcik2004quasiperiodic, buerschaper2004stroboscopic, vstefavnak2008recurrence, vstefavnak2008recurrencea, vstefavnak2009recurrence, vstefavnak2010full, grunbaum2013recurrence, cedzich2013propagation, bourgain2014quantum, friedman2017quantum, yin2019large, grunbaum2018generalization, grunbaum2020quantum}. In particular, it is worth noting that classical random walks, standard QWs, open QWs, and quantum Markov chains in general can be analyzed within the same mathematical framework \cite{grunbaum2018generalization, grunbaum2020quantum}.

To characterize the existence of recurrences in classical random walks, a quantity that refers to the probability that the walker returns to its initial position at any point of its evolution was introduced \cite{polya1921aufgabe}. This quantity is known as P\'olya number, and its use was extended to study recurrences in QWs \cite{vstefavnak2008recurrence, vstefavnak2008recurrencea}.

Recurrence of an arbitrary state is sometimes referred to as revival. Thus, one can consider full-state revivals of a QW, which means that both the coin system and the walker return to a given joint state. In fact, several studies have been conducted on state revivals in QWs \cite{dukes2014quantum, xue2015experimental, cedzich2016revivals, konno2017periodicity, jayakody2019full}. In particular, the conditions for a quantum walker on a cyclic path to exhibit state revival are presented in Ref. \cite{dukes2014quantum}. Also, two periodic state revivals in a single-photon one-dimensional QW governed by a time-dependent coin-flip operator were experimentally observed \cite{xue2015experimental}. Later, a theoretical explanation of QWs with quasiperiodic-time-dependent coin-flip operators, which includes the previous experiment, was presented \cite{cedzich2016revivals}. Moreover, it has been shown that the two-dimensional Grover walk (which is governed by a four-dimensional coin system) results in a two-step full-state revival \cite{vstefavnak2010full}. In contrast, the Hadamard walk in a cycle exhibits full-state revivals only for two, four, and eight-step periods \cite{konno2017periodicity}.

That said, it seems that full-state revivals in higher dimensional QWs have not been explored enough yet. Previous work has shown that there exists a (time-independent) coin-flip operator such that revivals with any desirable even period are a built-in feature of the resultant QW \cite{jayakody2019full}. It is important to point out that this result concerned a class of walks for which these revivals occur without any intervention.

In this article, however, we introduce an intervention protocol to induce revival on QWs that, generally speaking, would not present such property otherwise. More specifically, we prove that a large class of QWs on a $d$-dimensional lattice governed by a $c$-state coin-flip operator admits an intervention protocol that induces on-demand full-state revivals. The induced revivals correspond to recurrences of the joint state of the walker and coin system. With only two interventions on the coin system, these revivals are manifested after an arbitrary even number of steps. Furthermore, we extend the notion of P\'olya numbers to define a quantifier for revivals in QWs. We also introduce the \textit{partial P\'olya number} to characterize the first revival of the walk.

The paper is organized as follows. In Sec. \ref{sec:1d}, we develop our revival scheme for a QW on a line governed by a two-dimensional coin system and present some of its important properties. We also discuss these revivals in terms of partial P\'olya numbers. The generalization of this result to a QW on a $d$-dimensional lattice governed by a $c$-dimensional coin is given in Sec. \ref{sec:arbitrary}. The final discussion and outlook are given in Sec. \ref{sec:discussion}. Multiple appendices provide details on technical derivations of the results presented in the core portion of the article.

\section{QW on a one-dimensional lattice with a two-dimensional coin}
\label{sec:1d}

Consider the standard QW on a line, i.e., the evolution of a system (the walker) on a one-dimensional lattice conditioned on the state of a two-level system (the coin). The Hilbert space associated with the joint system will be denoted by $\mathcal{H}_c \otimes \mathcal{H}_w$, where $\mathcal{H}_c$ is the space of the coin and $\mathcal{H}_w$ is the space of the walker. A single-step progression of the system consists of a transformation applied to the coin system (i.e., the coin is tossed) followed by a conditional shift of the walker (the walker moves either to the left or right upon conditioning on coin outcome). We write the unitary operator corresponding to this single-step evolution as $U=SC$, where $S$ and $C$ are the shift and coin-flip operators, respectively. Here, we focus on the conventional shift operator, defined as
\begin{equation}
    S = \sum_{x\in\mathbb{Z}} \sum_{z\in\mathbb{Z}_2} |z\rangle\langle z| \otimes |x-(-1)^z\rangle \langle x|.
    \label{eq:1d-s}
\end{equation}
A general coin-flip operator of a two-dimensional coin system can be written as \cite{manouchehri2013physical}
\begin{equation}
    \begin{aligned}
        C = &\cos\theta |0\rangle\langle0| + e^{i\phi_1} \sin\theta |0\rangle\langle1| \\
        &+ e^{i\phi_2} \sin\theta |1\rangle\langle0| - e^{i(\phi_1+\phi_2)} \cos\theta |1\rangle\langle1|.
    \label{eq:1d-c}
    \end{aligned}
\end{equation}

Then, assuming the joint system starts in the state $|\Psi_0\rangle$, its state after $t$ steps is $|\Psi(t)\rangle = U^t |\Psi_0\rangle$. Now, we introduce an intervention on the coin degree of freedom consisting of applying a different coin-flip operator at specific time steps during the evolution of the joint system. For such an intervention, consider the ``coin-flip'' operator
\begin{equation}
    G = e^{i\phi_1} |0\rangle\langle1| - e^{i\phi_2} |1\rangle\langle0|.
\end{equation}
Also, write $V=SG$. Combining $U$ and $V$ accordingly, we model a scheme of QW in which interventions on the coin space are introduced at specific time steps as
\begin{equation}
    |\Psi(\tau)\rangle = (W_l)^r |\Psi_0\rangle,
    \label{eq:qw-scheme}
\end{equation}
where $W_l=U^lV$, $\tau =r(l+1)$ and $l,r\in\mathbb{N}$.

However, our choice for the coin-flip operators $C$ and $G$, together with the symmetry of the walk imposed by the shift operator $S$, gives a very special property to $W_l$. In fact, observe that
\begin{equation}
    G^\dag CG^\dag = C^\dag.
    \label{eq:prop1}
\end{equation}
and $G^2 = - e^{i(\phi_1+\phi_2)} I$. Moreover, $G$ has important relations to the topology of the walk, which is manifested through the identities $G^\dag SG = S^\dag$ and $GSG^\dag = S^\dag$. As a result,
\begin{equation}
    W_l = e^{i\Phi(l+1)} W_l^\dag,
    \label{eq:rel-1d-qw}
\end{equation}
where $\Phi = \phi_1+\phi_2 + \pi$. This is proven in Appendix \ref{app:1d-wl}. It means that $W_l$ is almost self-adjoint. In fact, if $\Phi=2\pi n/(l+1)$ for some integer $n$, the unitary $W_l$ is self-adjoint.

The above characteristic of $W_l$ is what leads to the revival property we want to obtain with the coin interventions. This is the case because it implies that
\begin{equation}
    W_l^2 = e^{i\Phi(l+1)} I.
    \label{eq:identity1d}
\end{equation}
Note that to backpedal the QW to its initial state, we need to apply the operator $V$ two times during a single cycle. This is the case regardless of the initial state of the system.

\begin{figure*}
    \centering
    \includegraphics[width=\textwidth]{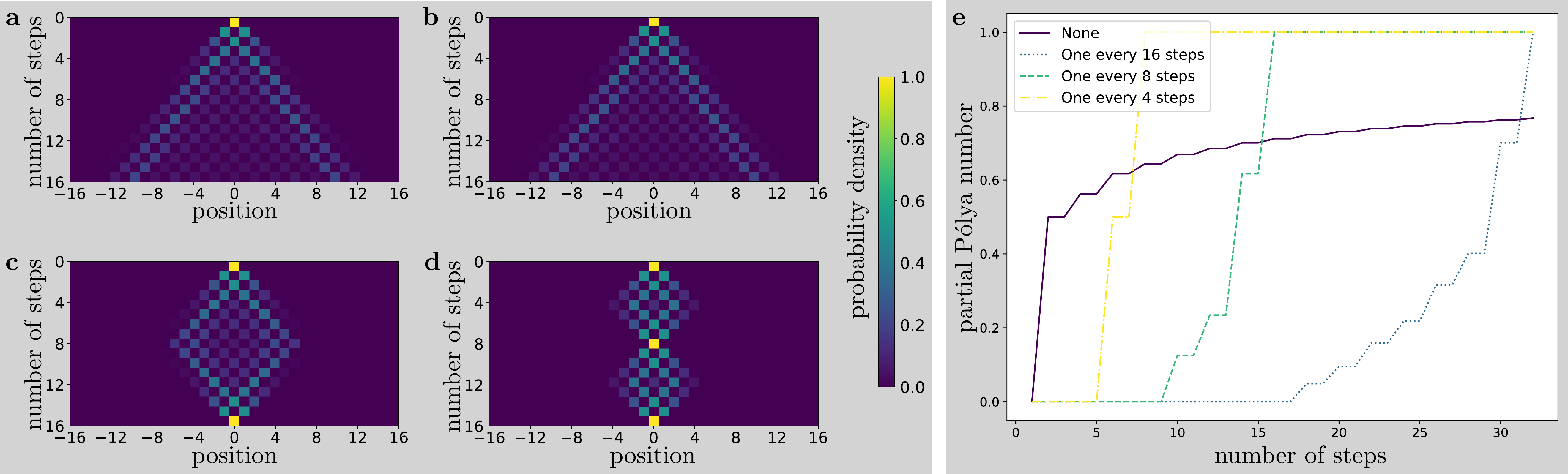}
    \caption{\textbf{Revival in a one-dimensional quantum lattice with a two-dimensional coin.} With periodic interventions given by a specially designed coin-flip operator $G$, revival is induced in a quantum walk system with an arbitrary coin-flip operator $C$. This is the case because the intervention with $G$ in the first step followed by $l$ steps with the regular coin-flip of the walk, which characterizes the unitary $W_l$, is, up to a phase, equivalent to the reverse process in time, i.e., $W_l^\dag$. \textbf{(a--d)} Probability distribution of a Hadamard walk with a walker initially localized at $|0\rangle$ and a coin started in the state $(|0\rangle + i|1\rangle)/\sqrt{2}$. Positions on the lattice are shown in the $x$ axis, while the $y$ axis displays the evolution in time. \textbf{(a)} Walk with no intervention, i.e., only the Hadamard coin-flip $C$ is used. \textbf{(b)} A single intervention in the first step with $G=|0\rangle \langle1| - |1\rangle \langle0|$ is made. Because of the initial state of the coin, no change to the probability distribution of the system is observed. \textbf{(c)} Two interventions made in the displayed time window, which characterizes $W_{7}$. \textbf{(d)} Interventions characterized by $W_{3}$ -- a total of four in the displayed time window. \textbf{(e)} Partial P\'olya number analysis of the walk with different periodicities of intervention. The quantity becomes one upon achievement of the first full-state revival.}
    \label{fig:cycle}
\end{figure*}

For illustration purposes, consider the example of a Hadamard walk with the walker starting localized at $|0\rangle$ and the initial state of the coin being $(|0\rangle + i|1\rangle)/\sqrt{2}$. The evolution of the probability distribution of this walk is represented in Fig. \ref{fig:cycle}a. Now, suppose the coin-flip operator $G=|0\rangle \langle1| - |1\rangle \langle0|$ is used before the first step of the walker instead of the Hadamard coin-flip $C$, as is the case of Fig. \ref{fig:cycle}b. Then, as can be seen, the probability distribution of the walker at later times is not affected by this initial replacement of the coin-flip operator. This is a special case due to the initial state of the coin. Interestingly, however, if $G$ is again used after $l$ applications of $C$, the walk will present revival after $l$ other coin-flips given by $C$. The time evolution of particular examples of periodic interventions associated with $W_7$ and $W_3$ are displayed in Fig. \ref{fig:cycle}c and Fig. \ref{fig:cycle}d, respectively.

In Fig. \ref{fig:cycle}e, we provide an alternative way to observe how fast the first revival of the walk takes place. For that, we start from the idea of a P\'olya number \cite{polya1921aufgabe}. While this quantity was originally associated with the study of recurrences on classical random walks, its use was extended to QWs in Refs. \cite{vstefavnak2008recurrence, vstefavnak2008recurrencea}. However, as it is explained in Appendix \ref{app:polya}, this concept can be further generalized to study full-state revival on QWs. In this case, the generalized P\'olya number becomes
\begin{equation}
    P_{FSR} \equiv 1 - \Pi_{t=1}^\infty \{1-\text{tr}[\pi_{\Psi(t)} \pi_{\Psi_0}]\},
    \label{eq:def-pnumber-full-rev}
\end{equation}
where $\pi_{\Psi} = |\Psi\rangle \langle\Psi|$, $|\Psi_0\rangle$ is the initial state of the walk, i.e., the initial joint state of the walker and the coin system, and $|\Psi(t)\rangle$ is the state of the walk at instant $t$. Observe that this quantity becomes one if a revival occurs at any step of the walk. However, the quantity displayed in Fig. \ref{fig:cycle}e corresponds to partial P\'olya numbers
\begin{equation}
    P_{FSR}^{(n)} \equiv 1 - \Pi_{t=1}^n \{1-\text{tr}[\pi_{\Psi(t)} \pi_{\Psi_0}]\}.
    \label{eq:def-partial-polya}
\end{equation}
The faster this quantity becomes one with the increase of $n$, the faster the walk presents revival, as explained in Appendix \ref{app:polya} and clearly seen in Fig. \ref{fig:cycle}e. Observe that, without intervention, the partial P\'olya number of the walk considered in Fig. \ref{fig:cycle} never reaches one, i.e., the walk does not presents revival.

\section{General case}
\label{sec:arbitrary}

We want to generalize the previous result to an arbitrary translationally-invariant QW on a lattice of any dimension $d$ where $d \in \mathbb{N}$. For that, like in Refs. \cite{vstefavnak2008recurrence, vstefavnak2008recurrencea}, assume $\{|\mathbf{m}\rangle\}$, where $\mathbf{m}\in\mathbb{Z}^d$, is a basis of $\mathcal{H}_w$ for which each element corresponds to the walker being at $\mathbf{m}$. Also, let $c$ be the number of possible translations of the walker after a single step, which corresponds to the dimension of $\mathcal{H}_c$. Denoting these translations by $\mathbf{e}_j$, the ``computational'' basis for the coin space can be written as $\{|\mathbf{e}_j\rangle\}$. Then, the shift operator can be written as
\begin{equation}
    S = \sum_{\mathbf{m}\in\mathbb{Z}^d} \sum_{j\in\mathbb{Z}_c} |\mathbf{e}_j\rangle \langle \mathbf{e}_j| \otimes |\mathbf{m} + \mathbf{e}_j\rangle \langle \mathbf{m}|.
\end{equation}
In addition to the translationally-invariance, we also assume that
\begin{equation}
    \mathbf{e}_j+\mathbf{e}_{c-1-j}=0
    \label{eq:symmetry}
\end{equation}
for every $j\in\mathbb{Z}_c$. In words, we assume that if the walker is allowed to walk a certain number of units in a given direction at each step, it must also be allowed to walk the same quantity of units in the opposite direction.

With the defined notation, the state of the system at any instant of time can be written as
\begin{equation}
    |\Psi(t)\rangle = \sum_{\mathbf{m}\in\mathbb{Z}^d} \sum_{j\in\mathbb{Z}_c} \psi_j(\mathbf{m}, t) |\mathbf{e}_j\rangle \otimes |\mathbf{m}\rangle.
\end{equation}

Moreover, it is possible to introduce the vector
\begin{equation}
    \psi(\mathbf{m},t) = (\psi_0(\mathbf{m},t), \psi_1(\mathbf{m},t), \cdots, \psi_{c-1}(\mathbf{m},t))^T.
\end{equation}
Then, if $C$ is the coin-flip operator, it follows that
\begin{equation}
    \psi(\mathbf{m},t) = \sum_{j\in\mathbb{Z}_c} C_j \psi(\mathbf{m}- \mathbf{e}_j,t-1),
    \label{eq:rec-rel-x}
\end{equation}
where the components of the matrices $C_j$ are given by $\langle\mathbf{e}_k|C_j|\mathbf{e}_l\rangle = \delta_{jk} \langle\mathbf{e}_k|C|\mathbf{e}_l\rangle$.

Because of the translational invariance we are assuming in this work, the recurrence relation in Eq. \eqref{eq:rec-rel-x} is independent of $\mathbf{m}$. Then, its analysis can be vastly simplified with the study of its Fourier transform $\tilde{\psi}(\mathbf{k},t)$, as suggested in Refs. \cite{vstefavnak2008recurrence, vstefavnak2008recurrencea}. In fact, the recurrence relation becomes
\begin{equation}
    \tilde{\psi}(\mathbf{k},t) = \tilde{C}(\mathbf{k}) \tilde{\psi}(\mathbf{k},t-1),
\end{equation}
where $\tilde{C}(\mathbf{k}) \equiv D(\mathbf{k}) C$ and $D(\mathbf{k}) \equiv \sum_{j\in\mathbb{Z}_c}e^{-i\mathbf{k}\cdot\mathbf{e}_j} |\mathbf{e}_j\rangle\langle\mathbf{e}_j|$. To give a more concrete example of this exposition, in Appendix \ref{app:fa-qw}, we provide the Fourier analysis of the one-dimensional QW considered in the previous section.

Like in the special one-dimensional case we just studied, we intend to introduce a second coin-flip operation $G$ in order to induce a state-independent revival on the QW. For that, there should exist at least one requirement over $G$ to establish how it is related to $C$ and the topology of the walk, similarly to the properties $G$ satisfied in the previous section. In this regard, it is shown in Appendix \ref{app:wlk-proof} that if
\begin{equation}
    \tilde{G}^\dag(\mathbf{k}) \tilde{C}(\mathbf{k}) \tilde{G}(\mathbf{k}) = e^{i\Phi}\tilde{C}^\dag(\mathbf{k}),
    \label{eq:single-req}
\end{equation}
then
\begin{equation}
    \tilde{W}_l^2(\mathbf{k}) = e^{i\Phi l} \tilde{G}^2(\mathbf{k}),
    \label{eq:gen-rec}
\end{equation}
where $\tilde{W}_l\equiv \tilde{C}^l\tilde{G}$. In Appendix \ref{app:gflip}, we show that an operator $G$ that satisfies Eq. \eqref{eq:single-req} for some $\Phi$ only exists if there exists a real $\Lambda$ such that
\begin{equation}
    \tilde{G}^2(\mathbf{k}) = e^{i\Lambda} I.
    \label{eq:constraint}
\end{equation}
More precisely, it is shown that
\begin{equation}
    G = \sum_{m\in\mathbb{Z}_c} e^{i\phi_m} |\mathbf{e}_m\rangle \langle\mathbf{e}_{c-1-m}|,
    \label{eq:g-form}
\end{equation}
where the phases $\phi_m$ satisfy $\phi_m + \phi_{c-1-m} = \Lambda$ for every $m\in\mathbb{Z}_c$. Moreover, Appendix \ref{app:gflip} proves that our protocol only works for walks whose coin-flip operator $C$ is such that
\begin{equation}
    C_{mn} = e^{i(\Phi + \phi_m - \phi_n)} \overline{C}_{c-1-n,c-1-m},
    \label{eq:aux1}
\end{equation}
where $C_{mn} \equiv \langle\mathbf{e}_m|C|\mathbf{e}_n\rangle$.

This means that, for the appropriate walks, revival is achieved with two interventions. In addition to showing these results, Appendix \ref{app:gflip} also presents two examples of higher-dimensional Hadamard and Grover walks that admit the type of intervention required in our protocol.

Finally, although we consider a protocol starting with an intervention, which implies that the walker and the coin system will return to their state when the intervention was first made, it is also possible to engineer a protocol that leads to the revival of the state of the joint system a certain number of steps prior to the first intervention. This is proved in Appendix \ref{app:intervention}.

\section{Discussion}
\label{sec:discussion}

In this work, we have introduced a procedure to generate on-demand state revival in QWs on a $d$-dimensional lattice governed by a $c$-dimensional coin system. With only two interventions in the coin degree of freedom, the joint system of the walker and the coin is backpedaled to their initial state. As a potential application of the proposed intervention scheme, one may introduce, for instance, a delay in the propagation of a given quantum state for a desired time period. This could be helpful e.g. when attempting to coordinate several correlated QWs, possibly as part of a quantum simulation scheme.

From a practical perspective, our scheme may simulate a periodically driven quantum system \cite{dittrich1998quantum}. Indeed, revival or recurrence of a given state in periodically driven systems is a well-known and helpful concept \cite{saif2001quantum}. With this in mind, we hope that the protocol we have just presented will be experimentally implemented soon. A potential platform for this involves optical setups, like the one used in Ref. \cite{nejadsattari2019experimental} to realize cyclic QWs.

A limitation of our procedure is that the type of intervention we require does not exist for every QW if $c>2$. Even though the class of walks for which our scheme is valid includes notable cases, like higher-dimensional Hadamard walks, a remaining open question concerns the existence of different protocols to induce revivals in a larger class of walks. For instance, if the two considered interventions were not required to be the same, would it be possible to extend the class of QWs for which on-demand revival is induced? Moreover, to relax even more the conditions assumed here, if a larger number of not necessarily repeated interventions is allowed, is there a protocol to induce revival on an arbitrary QW on a $d$-dimensional lattice? These are questions that deserve further investigation.

Finally, we have also considered the notion of P\'olya numbers. These quantities had previously been used to classify QWs as recurrent/transient. Here, we have extended this notion to categorize QWs that present state revival and, moreover, to determine the instant of first revival. We hope this notion could be fruitful in the future for studying revivals in additional types of quantum walks.

\acknowledgements{We wish to thank Christopher Cedzich and two anonymous referees for helpful comments. M.N.J. was supported by the President Scholarship and the BINA Scholarship at Bar-Ilan University. E.C. acknowledges support from the Israel Innovation Authority under projects 70002, 73795 from the Quantum Science and Technology Program of the Israeli Council of Higher Education, from the Pazy foundation and from the Ministry of Science and Technology. This work was supported by grant No. FQXi-RFP-CPW-2006 from the Foundational Questions Institute and Fetzer Franklin Fund, a donor advised fund of Silicon Valley Community Foundation.}

\bibliography{citations}

\onecolumngrid

\appendix

\section{Proof of the almost self-adjointness of $W_l$}
\label{app:1d-wl}

In this appendix, we show that Eq. \eqref{eq:rel-1d-qw} holds. This result follows from direct computation using the properties of $G$ mentioned in Sec. \ref{sec:1d}. In fact,
\begin{equation}
    \begin{aligned}
        W_l &= (SC)^l SG \\
            &= -e^{i(\phi_1+\phi_2)} (SC)^{l-1} S G G^\dag C G^\dag G^\dag S G \\
            &= -e^{i(\phi_1+\phi_2)} (SC)^{l-1} S G C^\dag S^\dag \\
            &= -e^{i(\phi_1+\phi_2)} W_{l-1} (SC)^\dag.
    \end{aligned}
\end{equation}
Iterations of this process lead to
\begin{equation}
    W_l = (-1)^{l-1} e^{i(\phi_1+\phi_2)(l-1)} W_1 [(SC)^\dag]^{l-1}.
\end{equation}
Moreover, since
\begin{equation}
    \begin{aligned}
        W_1 &= SCSG \\
            &= e^{2i(\phi_1+\phi_2)} G^\dag G S G^\dag G^\dag C G^\dag G^\dag SG \\
        &= e^{2i(\phi_1+\phi_2)} G^\dag S^\dag C^\dag S^\dag \\
        &= e^{2i(\phi_1+\phi_2)} W_1^\dag,
    \end{aligned}
\end{equation}
we conclude that
\begin{equation}
    \begin{aligned}
        W_l &= (-1)^{l+1} e^{i(\phi_1+\phi_2)(l+1)} W_1^\dag [(SC)^\dag]^{l-1} \\
            &= (-1)^{l+1} e^{i(\phi_1+\phi_2)(l+1)} W_l^\dag,
    \end{aligned}
\end{equation}
which completes the proof.

\section{P\'olya Number and state revival}
\label{app:polya}

Recurrence in a classical random walk is characterized by the P\'olya number \cite{polya1921aufgabe}, which can be written as
\begin{equation}
    P = 1 - \frac{1}{\sum_{t=1}^\infty p_0(t)},
\end{equation}
where $p_0(t)$ is the probability that the walker returns to the origin at time step $t$. If $P=1$, the walk is said to be \textit{recurrent}. Otherwise, i.e., if $0\leq P < 1$, the walk is called \textit{transient}. Observe that the walk is recurrent if and only if the series $\sum_{t=1}^\infty p_0(t)$ diverges. Moreover, the expression
\begin{equation}
    P = 1 - \Pi_{t=1}^\infty [1-p_0(t)]
    \label{eq:def-pnumber}
\end{equation}
can also be used as a definition for the P\'olya number as it provides the same criteria for the classification of the walk \cite{vstefavnak2008recurrencea}. With the above formula, this notion can be extended to the study of recurrence in QWs, as shown in Ref. \cite{vstefavnak2008recurrence, vstefavnak2008recurrencea}. If the walker starts at $|0\rangle$ and its state after $t$ steps is given by $\rho_W(t)$, the probability $p_0(t)$ can be written as $p_0(t) = \text{tr}[\rho_W(t) \pi_0]$, where $\pi_0 = |0\rangle\langle0|$. Hence,
\begin{equation}
    P = 1 - \Pi_{t=1}^\infty \{1-\text{tr}[\rho_W(t) \pi_0]\}.
\end{equation}

Here, however, we are interested in the study of revival in QWs. Nevertheless, the definition of P\'olya number can be easily adapted to a generic revival of the walker. In fact, since $|0\rangle$ is just an element of an arbitrary orthonormal basis of the space of the walker and any given state can also be seen as such, it follows that, if $|\psi_0\rangle$ is the initial state of the walker (which is typically assumed to be pure), the number
\begin{equation}
    P = 1 - \Pi_{t=1}^\infty \{1-\text{tr}[\rho_W(t) \pi_{\psi_0}]\},
    \label{eq:def-pnumber-rev}
\end{equation}
where $\pi_{\psi_0} = |\psi_0\rangle \langle\psi_0|$, equals one if and only if the walker presents ``recurrence to $|\psi_0\rangle$,'' i.e., if there exists a revival of $|\psi_0\rangle$.

Now, to include the state of the coin in the analysis and obtain a quantifier for the full-state revival, let $|\Psi_0\rangle$ and $|\Psi(t)\rangle$ be the initial state and the state at an instant $t$ of the system composed by the walker and the coin. Then, the quantity defined in Eq. \eqref{eq:def-pnumber-full-rev} quantifies the probability of full-state revival of the walk.

We can also define a partial P\'olya number for full-state revival as in Eq. \eqref{eq:def-partial-polya}. Observe that if the first full-state happens occurs deterministically in the $n$-th step, $P_{FSR}^{(s)}<1$ for every $s<n$ and $P_{FSR}^{(s)}=1$ for every $s\geq n$. Then, partial P\'olya number helps to determine the first full-state revival of the walk.

\section{Fourier analysis of the one-dimensional QW}
\label{app:fa-qw}

Here, for clarity, we present the Fourier analysis of the one-dimensional QW considered in Sec. \ref{sec:1d}. First, we write the state of the system at an instant of time $t$ as
\begin{equation}
    |\Psi(t)\rangle = \sum_{x\in\mathbb{Z}} \left[\psi_0(x, t) |0\rangle \otimes |x\rangle + \psi_1(x, t) |1\rangle \otimes |x\rangle\right].
\end{equation}
This allows us to introduce the vector
\begin{equation}
    \psi(x,t) = (\psi_0(x,t), \psi_1(x,t))^T.
\end{equation}
Then, if $C$ is the coin-flip operator and, hence, can written as in Eq. \eqref{eq:1d-c}, it follows that
\begin{equation}
    \psi(x,t) = C_0 \psi(x+1,t-1) + C_1 \psi(x-1,t-1),
    \label{eq:rec-rel-append}
\end{equation}
where $C_0 = \cos\theta |0\rangle\langle0| + e^{i\phi_1} \sin\theta |0\rangle\langle1|$ and $C_1 = e^{i\phi_2} \sin\theta |1\rangle\langle0| - e^{i(\phi_1+\phi_2)} \cos\theta |1\rangle\langle1|$.

With that, we are in position to use the translational invariance of the walk, which makes the recurrence in Eq. \eqref{eq:rec-rel-append} independent of $n$. As mentioned in Sec. \ref{sec:arbitrary}, this suggests a simplification of the analysis with the use of the Fourier transform $\tilde{\psi}(k,t)$ of $\psi(x,t)$. In fact, Eq. \eqref{eq:rec-rel-append} implies that
\begin{equation}
    \begin{aligned}
        \tilde{\psi}(k,t) &= e^{-ik} C_0 \tilde{\psi}(k,t-1) + e^{ik} C_1 \tilde{\psi}(k,t-1) \\
             &= \tilde{C}(k) \tilde{\psi}(k,t-1),
    \end{aligned}
\end{equation}
where $\tilde{C}(k) = D(k) C$ and $D(k) = e^{-ik} |0\rangle\langle0| + e^{ik} |1\rangle\langle1|$.

With that, Eq. \eqref{eq:qw-scheme} leads to
\begin{equation}
    \tilde{\psi}(k,m(l+1)) = \tilde{W}_l^d(k) \tilde{\psi}(k,0),
\end{equation}
where $\tilde{W}_l(k) = \tilde{C}^l(k) \tilde{G}(k)$ and $\tilde{G}(k) = D(k)G$.

To conclude, observe that the coin-flip operator of intervention $G$ defined in Sec. \ref{sec:1d} is such that
\begin{equation}
    \tilde{G}^2(k) = e^{i\Phi} I,
\end{equation}
where $\Phi=\phi_1+\phi_2+\pi$. Moreover, the requirement in Eq. \eqref{eq:single-req} is met. This can be easily checked since the above equation allows us to rewrite this requirement as $\tilde{G}^\dag(k) \tilde{C}(k) \tilde{G}^\dag(k) = \tilde{C}^\dag(k)$, which holds if and only if Eq. \eqref{eq:prop1} holds. Combined, these two properties lead to induced revival of the walk with two interventions, as is discussed in Sec. \ref{sec:arbitrary}.

\section{Proof of the main property of $\tilde{W}_l(\mathbf{k})$}
\label{app:wlk-proof}

The aim of this appendix is to show that if the identity in Eq. \eqref{eq:single-req} is satisfied, then Eq. \eqref{eq:gen-rec} holds. This follows from direct computation. In fact,
\begin{equation}
    \begin{aligned}
        \tilde{W}_l(\mathbf{k}) &= \tilde{C}^l(\mathbf{k}) \tilde{G}(\mathbf{k}) \\
              &= \tilde{C}^{l-1}(\mathbf{k}) \tilde{G}(\mathbf{k}) \tilde{G}^\dag(\mathbf{k})) \tilde{C}(\mathbf{k}) \tilde{G}(\mathbf{k}) \\
              &= e^{i\Phi} \tilde{C}^{l-1}(\mathbf{k}) \tilde{G}(\mathbf{k}) \tilde{C}^\dag(\mathbf{k}) \\
              &= e^{i\Phi} \tilde{W}_{l-1}(\mathbf{k}) \tilde{C}^\dag(\mathbf{k}),
    \end{aligned}
\end{equation}
which leads to
\begin{equation}
    \tilde{W}_l(\mathbf{k}) = e^{i\Phi(l-1)} \tilde{W}_1(\mathbf{k}) (\tilde{C}^\dag(\mathbf{k}))^{l-1}.
\end{equation}
Moreover, since
\begin{equation}
    \begin{aligned}
        \tilde{W}_1(\mathbf{k}) &= \tilde{C}(\mathbf{k}) \tilde{G}(\mathbf{k}) \\
              &= \tilde{G}(\mathbf{k}) \tilde{G}^\dag(\mathbf{k}) \tilde{C}(\mathbf{k}) \tilde{G}(\mathbf{k}) \\
              &= e^{i\Phi} \tilde{G}(\mathbf{k}) \tilde{C}^\dag(\mathbf{k}) \\
              &= e^{i\Phi} \tilde{G}^2(\mathbf{k}) \tilde{W}_1^\dag(\mathbf{k}),
    \end{aligned}
\end{equation}
we have
\begin{equation}
    \tilde{W}_l(\mathbf{k}) = e^{i\Phi l} \tilde{G}^2(\mathbf{k}) \tilde{W}_l^\dag(\mathbf{k}),
\end{equation}
which implies Eq. \eqref{eq:gen-rec}.

\section{Proof that $\tilde{G}^2(\mathbf{k})$ is proportional to the identity}
\label{app:gflip}

Writing a generic coin-clip $C$ as $C = \sum_{mn} c_{mn} e^{i\theta_{mn}} |\mathbf{e}_m\rangle \langle\mathbf{e}_n|$ and a generic $G$ as $G = \sum_{mn} g_{mn} e^{i\phi_{mn}} |\mathbf{e}_m\rangle \langle\mathbf{e}_n|$, the constraint in Eq. \eqref{eq:constraint} implies that, for every $m$ and $n$,
\begin{equation}
    \sum_s e^{i(-\mathbf{k}\cdot \mathbf{e}_s + \theta_{ms} + \phi_{sn})} c_{ms} g_{sn} = e^{i(\Phi + \mathbf{k}\cdot \mathbf{e}_n)} \sum_s e^{i(\phi_{ms} - \theta_{ns})} g_{ms} c_{ns}.
    \label{eq:terms-relation}
\end{equation}
Recalling the symmetry imposed in Eq. \eqref{eq:symmetry}, we observe that a consequence of the above expression is that
\begin{equation}
    c_{ms} g_{sn} = 0
\end{equation}
for every $m$ and $n$ and every $s\neq c-1-n$.

Now, suppose that there exists $G$ such that, for a given $n$ and $s\neq c-1-n$, we have $g_{sn}\neq0$. Then, the above relation implies that $c_{ms}=0$ for every $m$, i.e., $C$ contains a null column. Such a matrix cannot be a unitary. Therefore, the only non-null elements of $G$ are the ones in its secondary diagonal, i.e., $g_{n,c-1-n}$. As a result, we can simplify the notation and write Eq. \eqref{eq:g-form}.

With that, Eq. \eqref{eq:terms-relation} become
\begin{equation}
    C_{m,c-1-n} = e^{i(\Phi + \phi_m - \phi_{c-1-n})} \overline{C}_{n,c-1-m},
    \label{eq:adapted-relation}
\end{equation}
where $C_{mn}=e^{i(\theta_{mn})} c_{mn}$. On the one hand, by replacing $n\rightarrow c-1-n$, this allows us to write Eq. \eqref{eq:aux1}. On the other hand, taking the complex conjugate of Eq. \eqref{eq:adapted-relation} and replacing $m\rightarrow c-1-n$ and $n\rightarrow m$, we obtain
\begin{equation}
    C_{mn} = e^{i(\Phi + \phi_{c-1-n} - \phi_{c-1-m})} \overline{C}_{c-1-n,c-1-m}.
    \label{eq:aux2}
\end{equation}
Together, Eqs. \eqref{eq:aux1} and \eqref{eq:aux2} imply that $\phi_m + \phi_{c-1-m} = \phi_n + \phi_{c-1-n}$ for every $m$ and $n$. We, then, let $\Lambda$ be such that
\begin{equation}
    \Lambda \equiv \phi_m + \phi_{c-1-m}
    \label{eq:lambda-constraint}
\end{equation}
for every $m$. This means that if there exists $G$ that satisfies Eq. \eqref{eq:single-req}, it must be of the form given by Eq. \eqref{eq:g-form} with the constraint in Eq. \eqref{eq:lambda-constraint}. Moreover, a direct calculation shows that Eq. \eqref{eq:constraint} holds.

Furthermore, Eqs. \eqref{eq:constraint} and \eqref{eq:constraint} imply that
\begin{equation}
    G^\dag C G^\dag = e^{i(\Phi-\Lambda)} C^\dag.
    \label{eq:prop-gen}  
\end{equation}
In particular, if $\Phi$ equals $\Lambda$, Eq. \eqref{eq:constraint} reduces the requirement in Eq. \eqref{eq:single-req} to the relation in Eq. \eqref{eq:prop1} satisfied by coin-flip operators in the case studied in Sec. \ref{sec:1d}.

Another conclusion drawn from the results in this appendix is that the protocol proposed in this work is only valid for symmetric walks whose coin-flip operator is characterized by a matrix $C$ that satisfies Eq. \eqref{eq:aux1} -- or Eq. \eqref{eq:aux2} since they are equivalent.

To illustrate, consider \textit{unbiased walks}, which constitute a subclass of symmetric walks whose coin-flip operators satisfy
\begin{equation}
    |C_{mn}| = \frac{1}{\sqrt{c}}.
\end{equation}
It is clear that a subset of these walks meets the condition in Eq. \eqref{eq:aux1}. This subset includes, for instance, Hadamard and Grover walks.

To give a more concrete example, in the case $c=4$, the Hadamard coin of a walk on a two-dimensional lattice is characterized by $C_{mn}=(-1)^{m\cdot n}/2$, where $m\cdot n$ denotes the bitwise dot product of the binary representation of the indexes $m$ and $n$. It can be checked by direct computation that this coin-flip operator satisfies Eq. \eqref{eq:aux1}. We also conclude that the intervention operator in Eq. \eqref{eq:g-form} is such that $\phi_0 = \phi_3$ and $\phi_0-\phi_1 = \phi_0-\phi_2 = \pi$. Then, up to a global phase $G$ can be written as
\begin{equation}
   G = \left(
   \begin{array}{cccc}
      0 &  0 &  0 & 1 \\
      0 &  0 & -1 & 0 \\
      0 & -1 &  0 & 0 \\
      1 &  0 &  0 & 0
   \end{array} \right).
\end{equation}

Another notable example, as already mentioned, consists of Grover walks. To illustrate it further, we consider the walk on a two-dimensional lattice with a Grover coin, which is given by $C_{mn} = 1/2 - \delta_{mn}$, where $\delta_{mn}$ is the Kronecker delta. Direct calculation also shows that this operator satisfies Eq. \eqref{eq:aux1}. Moreover, it leads to an intervention operator of the form shown in Eq. \eqref{eq:g-form} such $\phi_m = \phi_n$ for every $m$ and $n$. Then, up to a global phase, we can write
\begin{equation}
   G = \left(
   \begin{array}{cccc}
      0 & 0 & 0 & 1 \\
      0 & 0 & 1 & 0 \\
      0 & 1 & 0 & 0 \\
      1 & 0 & 0 & 0 
   \end{array} \right).
\end{equation}

\section{Coin interventions in arbitrary time steps}
\label{app:intervention}

Here, we show that a protocol characterized by
\begin{equation}
    \tilde{Z}_{l,m}(\mathbf{k}) = \tilde{C}^{l-m}(\mathbf{k}) \tilde{G}(\mathbf{k}) \tilde{C}^m(\mathbf{k}),
\end{equation}
where $0\leq m\leq l$, can also lead to revivals in a similar manner obtained with $\tilde{W}_l(\mathbf{k})$.

First, observe that
\begin{equation}
    \begin{aligned}
        \tilde{Z}_{l,m}(\mathbf{k}) &= \tilde{C}^{l-m-1}(\mathbf{k}) \tilde{G}(\mathbf{k}) \tilde{G}^\dag(\mathbf{k}) \tilde{C}(\mathbf{k}) \tilde{G}(\mathbf{k}) \tilde{C}(\mathbf{k}) \tilde{C}^{m-1}(\mathbf{k}) \\
             &= e^{i\Phi} \tilde{C}^{l-m-1}(\mathbf{k}) \tilde{G}(\mathbf{k}) \tilde{C}^{m-1}(\mathbf{k}),
    \end{aligned}
    \label{eq:zw}
\end{equation}
where Eq. \eqref{eq:single-req} was used.

On the one hand, if $m\leq l/2$, repeated iterations of this process lead to
\begin{equation}
    \tilde{Z}_{l,m}(\mathbf{k}) = e^{i\Phi m} \tilde{C}^{l-2m}(\mathbf{k}) \tilde{G}(\mathbf{k}) = e^{i\Phi m} \tilde{W}_{l-2m}(\mathbf{k}).
\end{equation}
Hence, with Eq. \eqref{eq:gen-rec}, we conclude that
\begin{equation}
    \tilde{Z}_{l,m}^2(\mathbf{k}) = e^{2i\Phi m} \tilde{W}_{l-2m}^2(\mathbf{k}) = e^{i\Phi l} \tilde{G}^2(\mathbf{k}).
\end{equation}

On the other hand, if $m> l/2$, repeated iterations of Eq. \eqref{eq:zw} allow us to write
\begin{equation}
    \begin{aligned}
        \tilde{Z}_{l,m}(\mathbf{k}) &= e^{i\Phi (l-m)} \tilde{G}(\mathbf{k}) \tilde{C}^{2m-l}(\mathbf{k}) \\
             &= e^{i\Phi (l-m)} \tilde{G}(\mathbf{k}) \tilde{C}^{2m-l}(\mathbf{k}) \tilde{G}(\mathbf{k}) \tilde{G}^\dag(\mathbf{k}) \\
             &= e^{i\Phi (l-m)} \tilde{G}(\mathbf{k}) \tilde{W}_{2m-l}(\mathbf{k}) \tilde{G}^\dag(\mathbf{k}).
    \end{aligned}
\end{equation}
Therefore, using Eq. \eqref{eq:gen-rec}, we obtain
\begin{equation}
    \tilde{Z}_{l,m}^2(\mathbf{k}) = e^{2i\Phi (l-m)} \tilde{G}(\mathbf{k}) \tilde{W}_{2m-l}^2(\mathbf{k}) \tilde{G}^\dag(\mathbf{k}) = e^{i\Phi l} \tilde{G}^2(\mathbf{k}).
\end{equation}

\end{document}